\documentclass[twocolumn,preprintnumbers,nofootinbib]{revtex4-1}
\usepackage{amssymb}
\usepackage{mathrsfs}
\usepackage{float}
\usepackage{bm}
\usepackage{amsmath}
\usepackage{mathdots}
\usepackage{graphicx}
\usepackage{subfigure}
\usepackage{array}
\usepackage{lipsum}

\begin{document}
\newcommand*{\PKU}{School of Physics, Peking University, Beijing 100871,
China}\affiliation{\PKU}

\title{Sommerfeld effect as the vertex correction in three-dimensional space}

\author{Zhentao Zhang}\affiliation{\PKU}

\begin{abstract}
 It is shown that the correction to the vertex for fermion pair annihilation and production in the low-energy region is equal to the value of the wave function for the two-body system at the origin. The procedure also shows directly that the emergence of the Sommerfeld effect in quantum mechanics is the product of the reduction of the vertex correction from four spacetime dimensions to three-dimensional space. Meanwhile, the reciprocal wave function is introduced into quantum mechanics.
\end{abstract}
\maketitle

The Sommerfeld effect is an elementary effect in quantum mechanics~\cite{Sommerfeld}. It tells us that due to the nonperturbative interactions between the particles in the initial state, the wave function for the two-body system can be strongly distorted and that we should take into account the effect of this distortion in calculating the annihilation cross section. After considering this mechanism, the cross section may be written as $\sigma=S\sigma_0$, where $\sigma_0$ is the unperturbed cross section when there are no forces in the initial state, and the Sommerfeld factor $S$ is  phenomenologically defined as
\begin{equation}\label{S}
S=\frac{|\psi(0)|^2}{|\psi_0(0)|^2}=|\psi(0)|^2,
\end{equation}
where $\psi(r)$ is the scattering state wave function for the two-body system, and $\psi_0(r)$ is the plane wave. It is readily understood that the Sommerfeld factor is the probability of finding the particles at the origin relative to the case of no forces. Understandably, the Sommerfeld effect plays a broad role in particle physics research~\cite{Appelquist,Atwood,Cirelli,March-Russell,Arkani-Hamed,Lattanzi}.

At the micro-scale, the theory of quantum field offers us a more fundamental theoretical structure for describing our world, and thus we naturally desire that the mechanism for the nonperturbative phenomenon can be described within the structure of quantum field theory (QFT)~\footnote{I understand that if we adhere to reductionism for an elementary effect in quantum mechanics, practically we have only a very slim chance of being satisfied.}. The motivation can be even stronger, if one realizes the \textit{internal disharmony} that this quantum mechanics characterized factor is applied to calculating the quantum field theoretical cross section. The purpose of this paper is to establish the self-contained QFT mechanism and also the possible connection between the abstract structures of quantum mechanics and quantum field theory for the nonperturbative phenomenon.

It should be noted that to study the effect of the top quark width in its threshold pair production, a Green function method was introduced by Fadin and Khoze~\cite{Fadin}. The method was then improved by Strassler and Peskin through the Bethe-Salpeter equation in QFT, and the approach is associated with the optical theorem~\cite{Strassler}. Due to the optical theorem, one may find a threshold factor containing the imaginary part of the Green function in the approach if one calculates the cross section for the final state interactions above the threshold and then compares the final result with the unperturbed cross section, and however it would not unfold the basic quantum mechanical picture of the Sommerfeld effect.

Notice that for the Coulomb interactions, the Sommerfeld factor is
\begin{equation}\label{Coulomb}
S=\frac{\pi\alpha/v}{1-e^{-\pi\alpha/v}},
\end{equation}
where $\alpha$ is the Coulomb coupling constant, and $v$ is the velocity of the fermion in the center-of-mass frame. In quantum electrodynamics (QED), the threshold factor $1+\pi\alpha/(2v)$ was found by Schwinger in the one-loop perturbative calculation~\cite{Schwinger}. We know that in general the perturbative expansion would break down near the threshold. Thus, it can be understood that this threshold factor holds and agrees with the Sommerfeld factor in the perturbative zone $1\gg v \gg \alpha$.

For clarity, let us consider the nonperturbative mechanism in QED in this study.

\vspace{0.5cm}

\section*{The vertex correction}

From the elementary knowledge of the interactions in QED, one can understand that in principle the nonperturbative phenomenon must be somehow caused by the virtual photon correction for the vertex function.

\begin{figure}[H]
\centering
~~\includegraphics[width=8.5cm]{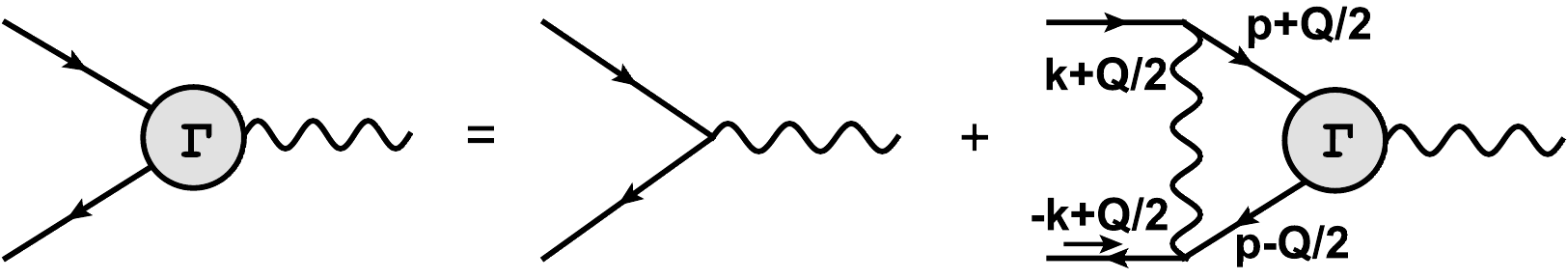}
 \caption{The Bethe-Salpeter equation for the vertex function in the ladder approximation. }\label{Ladder}
\end{figure}

The vertex function may be described by the Bethe-Salpeter equation, see Fig.~\ref{Ladder}.

The fermions in the diagrams are off-shell, i.e., $k_{1(2)}^2\neq m^2$~\cite{BLP}. In the center-of-mass frame, the variables $k$ and $Q$ are defined as
\begin{align}\label{variables}
  k&=\frac{1}{2}(k_1-k_2)=(k^0,{\bf k}),\\
  Q&=k_1+k_2=(E+2m,{\bf{0}}),
\end{align}
where $E+2m$ is the total energy, and $E(>0)$ is the energy of the relative motion.

In the integral equation for the low-energy system we need consider the contribution from the positive and the negative energy poles of the fermion propagators
\begin{align}
  S_{+}(p+Q/2)\rightarrow\frac{i}{2}\frac{1+\gamma^0}{E/2+p^0-{\bf{p}}^2/2m+i\epsilon},\\
  S_{-}(p-Q/2)\rightarrow\frac{i}{2}\frac{1-\gamma^0}{E/2-p^0-{\bf{p}}^2/2m+i\epsilon},
\end{align}
and we may then retain only the instantaneous part of the interactions from the photon exchange
\begin{equation}\label{instantaneous}
-\frac{\gamma^0\gamma^0}{({\bf{k-p}})^2}.
\end{equation}
Notice that there are two external legs of the low-energy fermions. Then the equation for the vertex function is
\begin{align}\label{field}
 \Gamma^\mu(k,Q)=\gamma^\mu+i&\int\frac{d^4p}{(2\pi)^4}[\frac{\Gamma^\mu(p,Q)}{(E/2+p^0-{\bf{p}}^2/2m+i\epsilon)} \nonumber \\
 &\times\frac{1}{(E/2-p^0-{\bf{p}}^2/2m+i\epsilon)}\frac{-e^2}{({\bf{k-p}})^2}].
\end{align}
From the dependence of the variables in the two sides of the equation, we may notice that $\Gamma^\mu(p,Q)$ should be independent of $p^0$. This is a natural consequence of the fact that the contribution from the noninstantaneous part of the interactions drops out in the nonrelativistic limit. Therefore, one can integrate the propagators over $p^0$.

Although it would be well-defined, from a ``more rigorous'' viewpoint, there is a problem for performing the integral immediately, since $\Gamma^\mu(p,Q)$ still contains the variable $p^0$ explicitly. To settle this purely formal, not physical, problem, we may inspect the structure of the equation. Recall that Lorentz invariance enforces us to introduce the relativistic normalization of the single particle state as
\begin{equation}
  \langle {\bf{q'|q}}\rangle=2q^0(2\pi)^3\delta^{(3)}({\bf{q-q}'}),
\end{equation}
where $q^0=\sqrt{m^2+{\bf q}^2}$. This normalization is different from the nonrelativistic normalization in quantum mechanics. And notice that comparing with the Born amplitude in quantum mechanics, for the scattering in the nonrelativistic domain we may need to redefine $\Gamma_4$ for canceling the contribution of the relativistic normalization factors in the amplitude, where $\Gamma_4$ is the scalar constituent of the relevant vertex function $\Gamma^\mu=\Gamma_4\gamma^\mu$. Clearly, the vertex function in Eq.~(\ref{field}) is under the relativistic normalization. Thus, it is natural to remove the pseudo momentum-dependence of the vertex function $\Gamma^\mu(p,Q)$ for the 0-component in the low-energy limit through the construction
\begin{equation}\label{remove}
\Gamma_4(p,Q)=\frac{\widetilde{\Gamma}({\bf{p}},E)}{2m},
\end{equation}
where $\widetilde{\Gamma}({\bf{p}},E)$ is the off-shell vertex function, nonrelativistically, depending on the variable in three-dimensional space, and the relativistic normalization factor for the two-body system is $2m$ in the nonrelativistic limit. Here $E$ occurs only as a parameter.

Thus, we have
\begin{align}
 \frac{\widetilde{\Gamma}({\bf{k}},E)}{2m}=\frac{1}{2m}+\int\frac{d^3p}{(2\pi)^3}\frac{m\widetilde{\Gamma}({\bf {p}},E)}{2m(mE-{\bf{p}}^2)}\frac{-e^2}{({\bf{k-p})}^2},
\end{align}
or
\begin{align}\label{new}
\widetilde{\Gamma}({\bf{k}},E)=1+\int\frac{d^3p}{(2\pi)^3}\frac{m\widetilde{\Gamma}({\bf{p}},E)}{(mE-{\bf{p}}^2)}\widetilde{V}({\bf{k-p})},
\end{align}
where $\widetilde{V}=-4\pi\alpha/({\bf{k-p}})^2$ is the Coulomb potential.

For the off-shell vertex function, let us set
\begin{equation}\label{int}
\widetilde{\Gamma}({\bf{k}},E)=(mE-{\bf{k}}^2)\widetilde{G}(E;{\bf{k}}),
\end{equation}
where $\widetilde{G}$ is an unknown function.

Using the Fourier transform, we may write Eq.~(\ref{new}) in the coordinate representation
\begin{equation}\label{new-coordi}
[{\nabla}^2-mV({\bf r})+mE]{G}(E;{\bf{r}})=\delta^{(3)}({\bf{r}}).
\end{equation}

Notice that, actually, any function of the continuous spectrum does not have its mathematical Fourier transform, since it would not be the (square-)integrable function. However, Dirac invented the bra-ket space in his representation theory, which is, formally, more general than a Hilbert space. As a consequence, the physical nonintegrable functions have their ``Fourier transforms'' in the sense of the representation transform. [It is important to notice that fundamentally the existence of this ``Fourier transform'' for giving the equivalent descriptions of the same physical reality in the different configuration spaces is not provided by the rigorously formal constructions in mathematics, but by the canonical commutation relation for the position and the momentum operators.]

In physics, it may also be useful to introduce the ``Fourier transform'' for the reciprocal form of the function of the continuous spectrum, which is also nonintegrable [say the reciprocal of the wave function $1/\varphi({\bf r})$]~\cite{Note1}. As we shall see later, it is now more appropriate to specifically regard the transform as the representation transform, and it should be

\begin{equation}\label{Definition}
\left.
\begin{aligned}
\mathcal{F}{\varphi^{-1}}&=(\mathcal{F}{\varphi})^{-1}={ {\chi^{-1}}},\\
\mathcal{F}^{-1}{\chi^{-1}}&=(\mathcal{F}^{-1}{\chi})^{-1}={ {\varphi^{-1}}},
\end{aligned}
\right\}
\end{equation}
where $\mathcal{F}$ and $\mathcal{F}^{-1}$ denote the operations of the Fourier transform, and $\chi$ is the wave function in the momentum representation. We may note for the time being that if the abstract quantity $\varphi^{-1}$ (or $\chi^{-1}$) itself is considered as a ``wave function'', the meaning of this reciprocal wave function might not be very transparent for us.

However, in certain circumstances, the reciprocal of the wave function may be the component in a well-defined physical quantity involving some abstract operations and/or the other abstract quantities. In this sense we can safely say that the reciprocal wave function contains the information of physics, and this kind of physical input may be called reciprocal contribution.

It is now useful to show that the transform for the reciprocal wave functions in Eq.~(\ref{Definition}) may have the same nature of the usual Fourier transform, i.e., it may be understood as a representation transform. To do that, we can make use of a reciprocal operation $\mathcal{D}$ which is a kind of operation that can be applied to any algebraic structure and does not rely on a particular configuration space, like the operation of the complex conjugate. When we apply this operation to the reciprocal wave function, the abstract quantity $\mathcal{D}\varphi^{-1}=1/\varphi^{-1}$ (or $\mathcal{D}\chi^{-1}$) has the clearly physical meaning, and there is
\begin{equation}\label{equivalence}
 1/{\varphi^{-1}}\Leftrightarrow1/{\chi^{-1}},
\end{equation}
where the symbol $\Leftrightarrow$ denotes the equivalence for describing the same physical state in the different representations ``through'' the Fourier transform. Since the reciprocal operation is not a physical operation, the relevant physical information in the wave function is formally stored in the algebraic structure of the reciprocal wave function. Thus we may have the correspondence
\begin{equation}\label{half-equivalence}
\varphi^{-1}\leftrightarrow\chi^{-1},
\end{equation}
where the symbol $\leftrightarrow$ denotes the equivalence of the reciprocal wave functions for describing the same physical reality in the coordinate and the momentum representations. And conversely, if we assume the correspondence for the reciprocal wave functions first, we may use the same operation to argue there is the correspondence for the wave functions. This reciprocity implies the  correspondences may be considered as the counterparts of each other in physics. Therefore, if a physical quantity involves the reciprocal contribution, for its representation transform, the transform for the reciprocal form may be employed~\cite{Note2}.

Returning now to our problem. From Eq.~(\ref{new}), we can see that in the absence of the interactions, $\widetilde{\Gamma}({\bf{k}},E)=1$. Thus, we have
\begin{equation}\label{free}
 (mE-{\bf{k}}^2)\widetilde{G}^0(E;{\bf{k}})=1
\end{equation}
for the case of no forces.

Using Eqs.~(\ref{int}) and (\ref{free}), we can rewrite the off-shell vertex function as
\begin{equation}\label{ratio}
{\widetilde{\Gamma}({\bf{k}},E)}=\frac{\widetilde{G}(E;{\bf{k}})}{\widetilde{G}^0(E;{\bf{k}})}.
\end{equation}
Note that the function $G$ of the continuous energy satisfies the inhomogeneous equation, and it would not be the Schr\"{o}dinger wave function. However, it is easy to understand the reciprocal contribution in a general way.

Then, integrating $\widetilde{G}/\widetilde{G}^0$ over the momentum and taking the representation transform for the reciprocal form, we have
\begin{equation}\label{in-ratio}
\bar{\Gamma}(E)\equiv\int d^3r_1\frac{G(E;{\bf{r_1}})}{G^0(E;{\bf{r-r_1}})|_{{\bf r}=0}}.
\end{equation}
After integrating over the possible momentum distribution in the two-body system, this vertex function contains the complete information about the vertex correction~\cite{Note3}.

Notice that $G(E;{\bf{r}})$ is identical with the Green function $G(E;{\bf{r,r'}})$ for ${\bf{r'}}=0$, and the Green function may be expanded as
\begin{equation}\label{G-expansion}
 G(E;{\bf{r,r'}})=\frac{1}{4\pi}\sum^{\infty}_{l=0}(2l+1)P_l(\cos\theta)r^{-1}r'^{-1}G_l(k;r,r'),
\end{equation}
where $G_l(k;r,r')$ is the partial-wave Green function, $\theta$ is the angle between $\bf{r}$ and $\bf{r'}$, and the symbol $k$ here denotes $\sqrt{mE}$.

For $r'=0$, only the $s$-wave contributes to the Green function. Notice $|{\bf{r-r_1}}|_{{\bf r}=0}=r_1$ and $P_0=1$. We have
\begin{equation}\label{G}
\bar{\Gamma}(E)=4\pi\int^\infty_0 r^2_1 d{r_1}{\frac{{G}_0(k;r_1,r'=0)}{{G^0_0}(k;r_1,r'=0)} },
\end{equation}
where ${G^0_0(k;r,r')}$ denotes the free $s$-wave Green function, and $G_0(k;r,r')$ satisfies the inhomogeneous equation
\begin{equation}\label{s-wave}
 [\frac{\partial^2}{\partial r^2}-mV(r)+k^2]G_0(k;r,r')=\delta(r-r').
\end{equation}

To construct a solution for Eq.~(\ref{s-wave}), we may use the regular and irregular solutions that satisfy the homogeneous equation
\begin{equation}\label{s-schrodinger}
[\frac{d^2}{d r^2}-mV(r)+k^2]\phi(r)=0.
\end{equation}
The regular solution $\phi_1(r)$ for the equation is solved with the boundary conditions at $r=0$
\begin{align}\label{regular}
 \phi_1(r)=0, \\
 \frac{d }{dr}\phi_1(r)=1.
\end{align}
And the irregular solution is the function $\phi_2(r)$ solved with the boundary condition at the infinity
\begin{align}\label{irregular}
 \phi_2(r)\rightarrow e^{ikr}.
\end{align}
However, a few words should be given here. Notice that for the asymptotic behavior of the irregular Coulomb solution, we remove the logarithmic term $-i\alpha \ln(2kr)/(2v)$ in the exponent. The appearance of this term is due to the fact that the Coulomb potential is not well-behaved. It is well known that this term would cause the singularity in the definition of the scattering amplitude. In practical calculations this frustrating term is ignored as an innocuously divergent phase factor. This implies the logarithmic term caused by the massless photon does not contribute to the observable quantities. So here we assume that the photon has an infinitesimal mass, and then the irregular solution has the well-behaved asymptotic behavior.

Since the Green function $G_0(k;r,r')$ would vanish at $r=0$ (for fixed $r'$) and we require it contains only the outgoing wave at $r\rightarrow\infty$, the Green function is of the form
\begin{equation}\label{Greenf1}
G_0(k;r,r')=P\phi_1(r_{<})\phi_2(r_{>}),
\end{equation}
where $r_{<}=\text{min}(r,r')$, $r_{>}=\text{max}(r,r')$, and $P$ is an undetermined parameter.

By the discontinuity of $\partial G_0/\partial r$ at $r=r'$
\begin{equation}\label{discontinuity}
{\left. \frac{\partial }{\partial r}G_0(k;r,r') \right|}^{r=r'+0}_{r=r'-0}=1,
\end{equation}
one may find
\begin{equation}
 P=-\frac{1}{\phi_2(0)}.
\end{equation}

Clearly, for the integral in Eq.~(\ref{G}), the Green functions are the functions for $(r)~r_1>r'$. For the free case, the Green function reads
\begin{equation}\label{f-Green}
 G^0_0(k;r_1,r')=-\frac{\sin{kr'}e^{ikr_1}}{k}.
\end{equation}

It can be seen that although $1/G^0_0(k;r_1,r'=0)$ diverges, it does not cause any singularity in the integral. And when $r_1$ is large,
\begin{equation}
 \frac{{G}_0(k;r_1,r'=0)}{{G^0_0}(k;r_1,r'=0)}=\frac{1}{\phi_2(0)}.
\end{equation}

Thus, we get
\begin{equation}\label{integration}
4\pi\int^\infty_0 r^2_1d{r_1}{\frac{{G}_0(k;r_1,r'=0)}{{G^0_0}(k;r_1,r'=0)} }=\frac{\delta^{(3)}(0)+C}{\phi_2(0)},
\end{equation}
where $C$ is a finite constant, and $\delta^{(3)}(0)$ is used to denote the infinite constant $\int d^3r_1$. As noted before, we are not interested in the result itself. It should be compared with the result for the free case ($V\rightarrow0$)
\begin{equation}\label{freeinte}
4\pi\int^\infty_0 r^2_1d{r_1}{\frac{{G_0}(k;r_1,r'=0)}{{G^0_0}(k;r_1,r'=0)} }=\delta^{(3)}(0),
\end{equation}
and the relative change between them measures the correction to the vertex $\text{C}_{\Gamma}$ at a particular energy.

Notice that $\phi_2(0)$ is the Wronskian of the two solutions $\phi_2(r)$ and $\phi_1(r)$ evaluated at the origin, and this Wronskian is known as the Jost function in scattering theory. For $\phi_2(0)$,
\begin{equation}\label{identity}
\frac{1}{\phi_2(0)}=\left.\frac{\phi_{3}(r)}{kr}\right|_{r=0}=\psi(0),
\end{equation}
where $\phi_{3}(r)$ is the solution of Eq.~(\ref{s-schrodinger}) with the ordinarily $s$-wave physical boundary conditions.

Combining Eqs.~(\ref{G}),~(\ref{integration}), (\ref{freeinte}), and (\ref{identity}), we obtain the correction value
\begin{equation}\label{exp}
\text{C}_{\Gamma}=\psi(0).
\end{equation}
We arrive, therefore, at an important conclusion that the correction to the vertex in three-dimensional space is equal to the value of the wave function at the origin. It should be pointed out that the meaning of this equality is twofold. On the other hand, it also clearly reveals the nature for the phenomenological factor: the change of the probability density for the particles at the origin is the vertex correction [$\psi(0)=\text{C}_{\Gamma}$].

It is of interest to note that since the vertex correction is not involved in the specific annihilation or production process, we may divide the scattering amplitude $\mathcal{M}$ into
\begin{equation}\label{amplitude}
 \mathcal{M}=\text{C}_{\Gamma}\cdot \mathcal{M}_{\text{unperturbed}},
\end{equation}
i.e., finding the correction is independent of the calculation of the unperturbed cross section. Recall that this is also the basic property of the Sommerfeld effect in quantum mechanics. Therefore, as a notable byproduct, our procedure directly shows that the emergence of the Sommerfeld effect in quantum mechanics is the product of the reduction of the vertex correction from four spacetime dimensions to three-dimensional space in the low-energy region.

\vspace{0.5cm}

\section*{Concluding remarks}

In this paper we calculate the vertex correction for fermion pair annihilation and production in the nonperturbative region, and we find that the correction to the vertex is equal to the value of the wave function for the two-body system at the origin. The multiple meanings of this study are easily seen, although it is nearly impossible to completely separate them from each other. It establishes the self-contained mechanism for the nonperturbative phenomenon within the structure of quantum field theory. Correlatively, it establishes an elegant connection between the basic structures of quantum mechanics and quantum field theory, and also shows the origin of the emergence of the Sommerfeld effect.

\vspace{1cm}

\appendix

\section*{Appendix}

\section*{A}
It would be natural to show that the transform between the coordinate and the momentum representations for the reciprocal wave functions should be Eq.~(15) by means of representation theory. Here we shall demonstrate the assertion in two other ways.

\vspace{0.2cm}
\textbf{I}) Let $|\overline{\Psi}\rangle$ be the ket. It denotes the abstract ``reverse image'' of the standard ket $|\Psi\rangle$. In the coordinate and the momentum representations,
\begin{align}
 \langle{\bf r}|\overline{\Psi}\rangle&=\Psi^{-1}({\bf r}), \tag{A.1} \\
 \langle{\bf p}|\overline{\Psi}\rangle&=\widetilde{\Psi}^{-1}({\bf p}), \tag{A.2}
\end{align}
if $\langle{\bf r}|\Psi\rangle=\Psi({\bf r})$ and $\langle{\bf p}|\Psi\rangle=\widetilde{\Psi}({\bf p})$.

[Notice that in quantum mechanics the wave function itself only has the meaning that is postulated to be the probability amplitude for the state of a system.]

Remarks are given as follows:
\begin{itemize}

\item The reciprocal wave function can have a physical meaning.

     To understand this fact, we recall that in quantum mechanics the meaning of the absolute probability amplitude for the normalized wave function is changed to the relative probability amplitude for the continuous spectral wave function. Accordingly, the reciprocal wave function is physically meaningful, because the ratio of the absolute square of it at two different points in a configuration space can still determine the relative probability of finding the system.

\item The reciprocal wave function itself can be interpreted as a relative probability amplitude, and \textit{the square of the modulus of the reciprocal wave function determines the relative probability of \underline{not} finding the system at the corresponding point in the configuration space}.

\end{itemize}

Certainly, the relative probability amplitude of not finding the system at a point in the configuration space is a ``passive'' description for a state. But since the finding and the nonfinding are mutually exclusive, the ``passive'' description does not contain less information about the state of a system comparing with the ``active'' description, i.e., \textit{the two descriptions are not independent of each other}. We may examine the physical consistency of the descriptions of the relative probability amplitudes of finding and nonfinding. Suppose $\overline{\psi}(q)$ is the configuration space wave function which represents the relative non-finding probability amplitude for a state. It can be understood that when the value of the absolute square of the wave function is proportional to the non-finding probability at a point, this value should be in inverse proportion to the chance of finding the system at that point. Thus, the quantity $1/|\overline{\psi}(q)|^2$ can be used to characterize the relative probability of finding the system in the configuration space. Then the relative probability of finding the system at two different points $q_1$ and $q_2$ is $|1/\overline{\psi}(q_1)|^2/|1/\overline{\psi}(q_2)|^2=|\overline{\psi}(q_2)|^2/|\overline{\psi}(q_1)|^2$. This means $\overline{\psi}(q)\propto1/{\psi}(q)$. Therefore, we can use the reciprocal of the wave function to characterize the relative non-finding probability amplitude in configuration space.

Notice that the probability of not finding a system in a small interval $\triangle q$ in the $q$ configuration space should be inversely proportional to $\triangle q$. Therefore, the quantities like [$|\Psi^{-1}(q)|^2\triangle q$ and] $\langle\overline{\Psi}|\overline{\Psi}\rangle$ are not meaningful. We may also understand this fact from a more fundamental standpoint. Since the reciprocal wave function can describe a state in configuration space, the ket $|\overline{\Psi}\rangle$ then corresponds to the same state represented by the ket $|\Psi\rangle$. And it is important to observe that, without a specific representation, we cannot really make a distinction between the abstract kets $|\overline{\Psi}\rangle$ and $|\Psi\rangle$. However, when they are both the kets corresponding to the state, the abstract description for the dynamical evolution of the state should be unique. So, logically we cannot use the ket $|\overline{\Psi}\rangle$, following the procedure for the ket $|\Psi\rangle$, to develop other abstract formalism for the dynamical evolution, and then the naive constructions for the state ket like $\langle\overline{\Psi}|\overline{\Psi}\rangle$ can be meaningless. [The postulates of quantum mechanics are deeply involved with the concept of measurement, and thus a complete mathematical formalism of quantum mechanics must be developed from the ``active'' viewpoint. (We can then consider $\langle\overline{\Psi}|\overline{\Psi}\rangle$ just an illegal expression of $\langle\Psi|\Psi\rangle$.)]

For this ket, one operation we are naturally allowed to consider is the transform between the coordinate and the momentum representations. For any ket that can describe a state of the system in configuration space, the abstract meaning of the ket, of course, does not affect the completenesses of the base kets $|{\bf r}\rangle$ and $|{\bf p}\rangle$. This then leads to
\begin{align}
 &\langle{\bf r}|\overline{\Psi}\rangle=\int \frac{d^3p}{(2\pi)^3}\langle{\bf r}|{\bf p}\rangle\langle{{\bf p}}|\overline{\Psi}\rangle,\tag{A.3}\\
 &\langle{\bf p}|\overline{\Psi}\rangle=\int d^3r\langle{\bf p}|{\bf r}\rangle\langle{{\bf r}}|\overline{\Psi}\rangle.\tag{A.4}
\end{align}
[It is noted that the base ket and its completeness are the concepts for describing a state of the system, not for the ket $|\Psi\rangle$ which itself is also an abstract concept corresponding to the state.]

We may write the transform in the algebraic forms
\begin{align}
&\Psi^{-1}({\bf r})\doteq\int \frac{d^3p}{(2\pi)^3} e^{i{\bf{p\cdot r}}}\widetilde{\Psi}^{-1}({\bf p}),\tag{A.5} \\
&\widetilde{\Psi}^{-1}({\bf p})\doteq\int d^3r e^{-i{\bf{p\cdot r}}}\Psi^{-1}({\bf r}),\tag{A.6}
\end{align}
where the symbol $\doteq$ is used to stress that it is the representation transform.

It can be noticed that, for the algebraic formulae of the representation transform, there is a ``wrong relation'' of the dimensional behavior between the two sides. To understand the innocuousness of this ``wrong relation'', we recall that the value of the reciprocal wave function is meaningful only in the relative sense, and that the wave function can always be multiplied by an arbitrary nonzero constant without changing its meaning. This means we can only require $\langle{\bf r}|\overline{\Psi}\rangle\propto{\Psi}^{-1}({\bf r})$ and $\langle{\bf p}|\overline{\Psi}\rangle\propto\widetilde{\Psi}^{-1}({\bf p})$. Therefore, for instance, we can define
\begin{align}
\langle{\bf p}|\overline{\Psi}\rangle&=\widetilde{\Psi}^{-1}_N({\bf p})=N\widetilde{\Psi}^{-1}({\bf p}),\tag{A.2a}
\end{align}
where the dimension of the constant $[N]=1/[m^{6}]$, and then the so-called ``wrong relation'' disappears. This dimensional term is trivial in physics. For the reciprocal wave functions, without affecting the physical meaning, we can retain the more algebraic symmetrical forms.

\vspace{0.2cm}

\textbf{II}) Since we have illuminated the physical meaning of the ket $|\overline{\Psi}\rangle$, we may also talk about the energy eigenvalue for this energy eigenket. However, if we want to get the energy eigenvalue from the ket, the generalization for the action of the operator on the ``normal'' ket is needed [In other words, we need to distinguish the ket $|\overline{\Psi}\rangle$ from $|\Psi\rangle$ by generalizing the measurement postulate of the observables for the ket]. For instance, if we apply the Hamiltonian operator $\hat{H}$ to the ket, to find its eigenvalue, it has to be understood as
\begin{equation}
 \hat{H}\rightarrow D\hat{H}D^{-1},\tag{A.7}
\end{equation}
where $D$ is the abstract mirror operation for the state ket, i.e., $D|\Psi\rangle=|\overline{\Psi}\rangle$ and $D^{-1}=D$. The leftmost $D$ is necessary for retaining the multiplication operations of the observables when they are applied to the ket.

Then
\begin{equation}
D\hat{H}D^{-1}|\overline{\Psi}\rangle=DE|\Psi\rangle=DED^{-1}|\overline{\Psi}\rangle,\tag{A.8}
\end{equation}
where $E$ is the energy eigenvalue of the state.

Correspondingly, when we apply the Fourier transform to the reciprocal wave functions, the operation may be understood as
\begin{align}
\mathcal{F}&\rightarrow D\mathcal{F}D^{-1}, \tag{A.9} \\
\mathcal{F}^{-1}&\rightarrow D\mathcal{F}^{-1}D^{-1}. \tag{A.10}
\end{align}
Notice that the Fourier transform is the map for the wave functions, and
\begin{align}
D^{-1}\langle{\bf r}|\overline{\Psi}\rangle&=\langle{\bf r}|D^{-1}|\overline{\Psi}\rangle, \tag{A.11}\\
D^{-1}\langle{\bf p}|\overline{\Psi}\rangle&=\langle{\bf p}|D^{-1}|\overline{\Psi}\rangle \tag{A.12}
\end{align}
can be well-defined since $D^{-1}$ acts on the state ket.

Eventually $D$ is considered, equivalently, to act on the configuration space wave function, and formally we have $D\langle{\bf r}|=\langle{\bf r}|D=\mathcal{D}\langle{\bf r}|$ and $D\langle{\bf p}|=\langle{\bf p}|D=\mathcal{D}\langle{\bf p}|$ [or $D\langle{\bf p}|=\mathcal{D}_{N}\langle{\bf p}|=N\mathcal{D}\langle{\bf p}|$ for the definition~(A.2a)], where $\mathcal{D}$ is the reciprocal operation in the representations. [In general $\hat{A}|~\rangle$ represents a new ket, and it is a constant times the original ket $|~\rangle$ only in the case that the operator $\hat{A}$ is an observable and $|~\rangle$ is the eigenket of it. Thus, for $D$, the abstract quantity like $E|\Psi\rangle[=\hat{H}|\Psi\rangle]$ should be considered as a ket (Notice that $D$ is an operation for the ket, not a physical operation for the state).]

The way of understanding the representation transform for the reciprocal wave function as a generalization of the Fourier transform is consistent with the preceding one, although it may not be as essential as the former and focuses on the formal aspects. This further development for the abstract concept of the ket $|\overline{\Psi}\rangle$ may have the conceptual advantages in giving the transformation a more general meaning, though from the consistency of the physical formalism, the former already implies that we can use the representation transform for any reciprocal of the physical nonintegrable function derived from the concept of the continuous spectral wave function.

\vspace{0.5cm}

At the end of this section, concerning the \textit{philosophy} of introducing the ket $|\overline{\Psi}\rangle$ from the ``passive'' viewpoint, we may quote the remark of R. P. Feynman~\cite{Feynman}: ``If we imagine some operator $\hat{A}$, we can use it with any state $|\psi \rangle$ to create a new state $\hat{A}|\psi \rangle$. Sometimes a `state' we get this way may be very peculiar$-$it may not represent any \textit{physical} situation we are likely to encounter in nature.'' ``In other words, we may at times get `states' that are mathematically artificial. Such artificial `states' may still be useful, perhaps as the mid-point of some calculation.''

\section*{B}
Besides removing the off-shell condition by integrating over the momentum, we may notice that there is the way for restoring the on-shell condition by setting ${\bf |k|}=\sqrt{mE}$ in the momentum representation. But remember that after finding the integral equation for the vertex function in ${\bf k}$-space, that we are able to transform the equation to ${\bf r}$-space depends upon the initial assumption that the fermions are off their mass-shells. Thus, it may not be self-consistent to consider the ``relation'' between the on-shell vertex function $\widetilde{\Gamma}_{\text{k}}\equiv{\widetilde{\Gamma}({\bf{k}},E)|_{{\bf |k|}=\sqrt{mE}}}$ and the function $G(E;\bf{r})$.

Nevertheless, in the optical theorem treatment for the final state interactions, the two ``external'' particles are the real off-shell particles in the diagram of the vacuum polarization, and then in this scenario it might be self-consistent to consider the relation between ${\widetilde{\Gamma}_{\text{k}}}$ and $G(E;\bf{r})$.

It is interesting to give a remark based on this point.

[In this treatment for the final state interactions the vertex function cannot be treated in isolation, and we simply have $\Gamma_4(k,Q)={\widetilde{\Gamma}({\bf{k}},E)}$.]

Using
\begin{equation}
 \text{Im}~\frac{1}{k-\sqrt{mE}-i0}=\pi\delta(k-\sqrt{mE}),     \tag{B.1}
\end{equation}
by Eq.~(13) (noticing also the scalar dependence of $\bf k$ for the $s$-wave contributed vertex function), we can find
\begin{equation}
\text{Im}~G(E;0)=-\frac{k}{4\pi}{\widetilde{\Gamma}_{\text{k}}}.\tag{B.2}
\end{equation}

The function $G(E;\bf{r})$ is identical with the Green function $G(E;{\bf{r,r'}}=0)$, and it can be shown that for the Green function that satisfies the outgoing wave boundary condition [which corresponds with the way of picking up the pole in Eq.~(B.1)], we have
\begin{equation}
 \text{Im}~G(E;0,0)=-\frac{k}{4\pi}|\psi(0)|^2.     \tag{B.3}
\end{equation}

By Eqs.~(B.2) and (B.3), we obtain
\begin{equation}
 {\widetilde{\Gamma}_{\text{k}}}=|\psi(0)|^2.      \tag{B.4}
\end{equation}
It may be emphasized, however, that this relation for the final state interactions needs to be understood under the specific physical pictures that i) the optical theorem treatment for the vacuum polarization; ii) the symmetry of the irreducible diagram for the photon self-energy, i.e., any correction caused by the photon exchange in the vacuum polarization can be assigned to the single vertex in the skeleton diagram.

\section*{C}

From an effective field theory viewpoint, it is very simple to consider (and calculate) the effective vertex correction for the nonpoint-type annihilation process, if the force carriers are much lighter than the particle which is responsible for the annihilation. And note that if we cannot consider the effective vertex correction in three-dimensional space for a nonpoint-type annihilation process, from the quantum mechanics point of view the Sommerfeld effect for the process also disappears, as it should be.

\section*{D}
In Ref.~\cite{Iengo}, it was concluded that the Sommerfeld factor was effectively obtained from field theory diagrams. Some equations used in the work were derived in its appendix B. But unfortunately the equations were found in the case of ``the first order in perturbation theory''. Notice that essentially the Sommerfeld effect is a nonperturbative effect, and that the perturbative expansion can break down in this scenario. Thus, it is incorrect to use the wave function of the Born approximation, the plane wave, in the derivation. Furthermore, for the Eq.~(2.13) in Ref.~\cite{Iengo}, the limit should be calculated after performing the integral over the momentum.

In addition, for understanding the singularities in the scattering state wave function in the momentum representation, we refer the reader to \S130 in L.~D.~Landau and E.~M.~Lifshitz, \textit{Quantum Mechanics}, Pergamon 1977.

\section*{E}
A treatment for the Sommerfeld effect in an effective way was given in Ref.~\cite{Cassel}. It was assumed that we may deal with the nonrelativistic two-body scattering first and that later we can put the wave function of the scattering state into the annihilation amplitude. Under this assumption, the scattering state was isolated and described by the inhomogeneous Bethe-Salpeter equation. Then the inhomogeneous term given by the single-boson exchange was neglected in the nonrelativistic limit, and the nonrelativistic wave function was found. While it is the familiar treatment for finding the bound state wave function, unfortunately it is illegal for the scattering state problem. Since generally there are no poles in the $S$-matrix (or the scattering amplitude), the inhomogeneous term cannot be dropped. And this is the major difference between the scattering-state and the bound-state Bethe-Salpeter equations.

\renewcommand{\thefootnote}{}

\footnote{Update: A follow-up paper is available on my personal website www.physzhang.com. Feel free to ask me for a copy when necessary.}

\end{document}